\documentclass[aps, prd, twocolumn]{revtex4-1}
\begin{document}
\title{Energy conditions in the epoch of galaxy formation}
\author{Matt Visser}
\address{Physics Department, Washington University, St. Louis, 
         Missouri 63130-4899}
\date{Originally submitted November 1996; Revised February 1997}
\begin{abstract}
\noindent
The energy conditions of Einstein gravity (classical general
relativity) do not require one to fix a specific equation of state.
In a Friedmann--Robertson--Walker universe where the equation of
state for the cosmological fluid is uncertain, the energy conditions
provide simple, model-independent,  and robust bounds on the behaviour
of the density and look-back time as a function of red-shift.
Current observations suggest that the ``strong energy condition''
is violated sometime between the epoch of galaxy formation
and the present.  This implies that no possible combination of
``normal''  matter is capable of fitting the observational data.

\bigskip
\noindent
Published in: Science {\bf276} (1997) 88-90;  doi: 10.1126/science.276.5309.88

\bigskip
\noindent
Currently at: School of Mathematics, Statistics, and Operations Research, \\
Victoria University of Wellington, PO Box 600, Wellington 6140,  New Zealand.
\end{abstract}
\maketitle
\def\implies{\Longrightarrow}
\def\sign{\hbox{sign}}

The energy conditions of Einstein gravity (classical general
relativity) place restrictions on the stress-energy tensor $T_{\mu\nu}$
(energy-momentum tensor)~\cite{Hawking-Ellis,Wald,Visser}. This
tensor is a 4 by 4 matrix built up out of the energy density,
momentum density, and the 3 by 3 stress tensor (pressure and
anistoropic stresses).  The energy conditions force various linear
combinations of these quantities to be positive and have been used,
for instance, to derive many theorems of classical general
relativity---such as the singularity theorems; the area increase
theorem for black holes, and the positive mass theorem---without
the need to assume a specific equation of
state~\cite{Hawking-Ellis2,Wald2,Visser2}. General relativists and
particle physicists would be surprised if large violations of the
classical energy conditions occur at temperatures significantly
below the Planck scale $k T < E_{Planck} \approx 10^{19}$ GeV,
$T\approx 10^{32}$ K~\cite{Violations}.  (Above the Planck scale
quantum gravity takes over, the whole framework of classical
cosmology seems to break down, and the question is
moot~\cite{Centenary,Three-hundred,Visser3}.)

Current observations seem to indicate that the strong energy
condition (SEC) is violated rather late in the life of the
universe---somewhere between galaxy formation and the present time,
in an epoch where the cosmological temperature never exceeds 60 K.
I shall show this by using the energy conditions to develop simple
and robust bounds for the density and look-back time~\cite{look-back}
as a function of red-shift in a Friedmann--Robertson--Walker (FRW)
cosmology.  The experimental observations I need are the present
day value of the Hubble parameter $H_0$, an age estimate for the
age of the oldest stars in the galactic halo, and an estimate for
the red-shift at which these oldest stars formed. From the theoretical
side, I only need to use a FRW cosmology subject to the Einstein
equations and classical energy conditions.  Using the energy
conditions to place bounds on physical parameters of the universe
allows me to avoid the need to separately analyze cold, hot,
lukewarm, or mixed dark matter.  Similarly,  MACHOS (massive compact
halo objects), WIMPS (weakly interacting massive particles), axions,
massive neutrinos, and other hypothetical contributions to the
cosmological density are automatically included as special cases
of this analysis.

The bounds I derive from the SEC are independent of whether or not
the universe is open, flat, or closed---which means that the density
parameter ($\Omega$ parameter, the ratio of the actual density to
the critical density needed to close the universe) does not have
to be specified. Thus my approach is independent of the existence
or nonexistence of any of the standard variants of cosmological
inflation~\cite{inflation}, which typically predict $\Omega =
1$~\cite{Kolb-Turner,Linde,Moss}.

If current observations are correct, then the ``strong energy
condition'' (SEC) must be violated sometime between the epoch of
galaxy formation and the present.  This implies that no possible
combination of ``normal''  matter is capable of fitting the
observational data, and one needs to do something drastic to the
cosmological fluid, either introduce a cosmological constant
$\Lambda$~\cite{Leonard-Lake}, or have a very non-standard weak
form of cosmological inflation that persists right up to galaxy
formation.


The spacetime geometry of the standard FRW cosmology is described
by specifying the geometry of space as a function of time, using
the spacetime metric~\cite{Peebles,Weinberg}

\begin{equation}
ds^2 = -dt^2 + a(t)^2 
\left[ {dr^2\over1-k r^2} + r^2(d\theta^2+\sin^2\theta d\phi^2)\right].
\end{equation}

\noindent
Here $ds$ is the  invariant interval between two events, $t$ is
comoving time (time as measured by an observer following the average
Hubble flow), $a(t)$ is the scale parameter describing the size of
the universe as a function of time, and $(r,\theta,\phi)$ are
spherical polar coordinates used to cover all of space (with space
at time $t$ being defined as the constant-$t$ slice through the
spacetime).  The parameter $k$ depends on the overall geometry of
space, and only takes on the values

\begin{equation}
k = 
\left\{
\matrix{+1              &    \hbox{closed} & (\hbox{if }\Omega > 1), \cr
        \hphantom{+} 0  &    \hbox{flat}   & (\hbox{if }\Omega = 1), \cr
        -1              &    \hbox{open}   & (\hbox{if }\Omega < 1). \cr}
\right.
\end{equation}

\noindent
The two non-trivial components of the Einstein equations yield the
total density $\rho$ and total pressure $p$ of the cosmological
fluid as a function of the scale factor $a$~\cite{CC-absorbed}.

\begin{equation}
\rho = {3\over8\pi G}
\left[{\dot a^2\over a^2} + {k\over a^2} \right].
\end{equation}

\begin{equation}
p = -{1\over8\pi G}
\left[2{\ddot a\over a} + {\dot a^2\over a^2} + {k\over a^2} \right].
\end{equation}

\noindent
They can be combined to deduce the conservation of stress-energy

\begin{equation}
\dot \rho = - 3 {\dot a\over a} (\rho + p).
\end{equation}

\noindent
Here $\dot a$ is the (time dependent) velocity of expansion of the
universe. Combined with the scale factor $a(t)$ it defines
the (time dependent) Hubble parameter

\begin{equation} 
H(t) = {\dot a(t) \over a(t)}.
\end{equation}

There are several different types of energy condition in general
relativity, the two main classes being averaged energy conditions
(that depend on some average of the stress-energy tensor along a
suitable curve), and the point-wise energy conditions (that depend
only on the stress-energy tensor at a given point in spacetime).
The standard point-wise energy conditions are the null energy
condition (NEC), weak energy condition (WEC), strong energy condition
(SEC), and dominant energy condition (DEC).  Basic definitions are
given  in~\cite{Hawking-Ellis,Wald,Visser} and for the special case of
a FRW spacetime the general formulae simplify.

\begin{equation}
\hbox{NEC} \iff \quad 
(\rho + p \geq 0 ).
\end{equation}

\begin{equation}
\hbox{WEC} \iff \quad 
(\rho \geq 0 ) \hbox{ and } (\rho + p \geq 0).
\end{equation}

\begin{equation}
\hbox{SEC} \iff \quad 
(\rho + 3 p \geq 0 ) \hbox{ and } (\rho + p \geq 0).
\end{equation}

\begin{equation}
\hbox{DEC} \iff \quad 
(\rho \geq 0 ) \hbox{ and } (\rho \pm p \geq 0).
\end{equation}

\noindent
The NEC is enough to guarantee that the density of the universe
goes down as its size increases.

\begin{equation}
\hbox{NEC} \iff \quad 
\sign(\dot \rho) = - \sign(\dot a).
\end{equation}

\noindent
If the NEC is violated the density of the universe must increase
as the universe expands---so something has gone very seriously
wrong.  The WEC additionally requires that the density is positive~\cite{wec}.

To understand what the SEC requires for the physical universe,
consider the quantity $d(\rho a^2)/dt$, and use the Einstein
equations to deduce

\begin{equation}
{d\over dt}(\rho a^2) = - a \dot a (\rho+3p).
\end{equation}

\noindent
Thus

\begin{equation}
\hbox{SEC} \implies \quad 
\sign \left[ {d\over dt}(\rho a^2) \right] = - \sign(\dot a).
\end{equation}

\noindent
This implies that 

\begin{equation}
\hbox{SEC} \implies \quad 
\rho(a) \geq \rho_0 (a_0/a)^2 \qquad \hbox{for} \qquad a< a_0.
\end{equation}

\noindent
In terms of the red-shift ($1+z= a_0/a$):

\begin{equation}
\hbox{SEC} \implies \quad 
\rho(z) \geq \rho_0 (1+z)^2.
\end{equation}

\noindent
The subscript zero denotes present day values, and the SEC  provides
a model-independent lower bound on the density of the universe
extrapolated back to the time of the big bang.  Another viewpoint
on the SEC comes from considering the quantity

\begin{equation}
\rho + 3 p = -{3\over4\pi G} \left[{\ddot a\over a} \right].
\end{equation}

\noindent
That is

\begin{equation}
\hbox{SEC} \implies \qquad \ddot a < 0.
\end{equation}

\noindent
The SEC implies that the expansion of the universe is decelerating---and
this conclusion holds independent of whether the universe is open,
flat, or closed.

For the DEC, use the Einstein equations to compute

\begin{equation}
{d\over dt}(\rho a^6) = + 3 a^5 \dot a (\rho-p).
\end{equation}

\noindent
Thus

\begin{equation}
\hbox{DEC} \implies \quad 
\sign\left[{d\over dt} (\rho a^6) \right] = + \sign(\dot a).
\end{equation}

\noindent
The DEC therefore provides an upper bound on the energy density.

\begin{equation}
\hbox{DEC} \implies \quad 
\rho(a) \leq \rho_0 (a_0/a)^6 \qquad \hbox{for} \qquad a< a_0.
\end{equation}

\noindent
In terms of the red-shift

\begin{equation}
\hbox{DEC} \implies \quad \rho(z) \leq \rho_0 (1+z)^6.
\end{equation}


When we look into the sky and see some object, the look-back time
to that object ($\tau= t_0-t$) is defined as the difference between
$t_0$  (the age of the universe now)  and  $t$ (the age of the
universe when the light that we are now receiving was emitted).
If we know the velocity of expansion of the universe $\dot a$ as
a function of scale parameter $a$ we simply have

\begin{equation}
\tau(a;a_0) = t_0-t = \int_a^{a_0} {da\over \dot a(a)}.
\end{equation}

\noindent
By putting a lower bound on $\dot a$ we deduce an upper bound on
look-back time. In particular since the SEC implies that the
expansion is decelerating then

\begin{equation}
\hbox{SEC} \implies \quad 
\tau(a;a_0) = t_0-t \leq {1\over H_0} {a_0-a \over a_0},
\end{equation}

\noindent 
independent of whether the universe is open,
flat, or closed.  Expressed in terms of the red-shift:

\begin{equation}
\hbox{SEC} \implies \quad 
\tau(z) = t_0-t \leq {1\over H_0} {z\over1+z}.
\end{equation}

\noindent
This provides us with a robust upper bound on the Hubble parameter

\begin{equation}
\hbox{SEC} \implies \quad 
\forall z: H_0 \leq {1\over \tau(z)} {z\over1+z}.
\end{equation}

This is enough to illustrate the age-of-the-oldest-stars problem
(often mischaracterized as the  age-of-the-universe problem).
Suppose there is some class of standard candles whose age of
formation, $\tau_f$, can be estimated~\cite{standard-candle}.
Suppose further that if we look out far enough we can see some of
these standard candles forming at red-shift $z_f$ (or can estimate
the red-shift at formation). Then

\begin{equation}
\hbox{SEC} \implies \quad H_0 \leq {1\over \tau_f} {z_f\over1+z_f} 
\leq {1\over \tau_f}.
\end{equation}

\noindent
The standard candles currently of most interest (simply because
they have the best available data and provide the strongest limit)
are the globular clusters in the halos of spiral galaxies: stellar
evolution models estimate (they do not measure)  the age of the
oldest stars still extant to be  $16\pm2 \times 10^9 \hbox{ yr}$
\cite{Peebles2}.  That is, at an absolute minimum

\begin{equation}
\hbox{Age of oldest stars} \equiv \tau_f 
\geq 16\pm2 \times 10^9  \hbox{ yr}.
\end{equation}

\noindent
Using $z_f < \infty$, this implies that~\cite{Megaparsec}

\begin{equation} 
H_0 \leq \tau_f^{-1} \leq 62\pm8 \hbox{ km s$^{-1}$ Mpc$^{-1}$ }.
\end{equation}

\noindent
When we actually look into the night sky, we infer that the oldest
stars seem to have formed somewhat earlier than the development of
galactic spiral structure~\cite{Peebles3}. A canonical first estimate
is~\cite{Peebles4}

\begin{equation}
\hbox{Redshift at formation of oldest stars}\equiv z_f \approx 15.
\end{equation}

\noindent
This now bounds the Hubble parameter

\begin{equation}
\hbox{SEC} \implies \quad 
H_0 \leq 58\pm7  \hbox{ km s$^{-1}$ Mpc$^{-1}$ }.
\end{equation}

\noindent
Recent estimates of the present day value of the Hubble
parameter are\cite{PDG}

\begin{equation} 
H_0 \in (65,85) \hbox{ km s$^{-1}$ Mpc$^{-1}$ }.
\end{equation}

\noindent
(I have chosen to use a range of $H_0$ values on which there is
widespread though not universal consensus~\cite{PDG}.) But even
the lowest reasonable value, $H_0=65$ km s$^{-1}$ Mpc$^{-1}$, is
only just barely compatible with the SEC, and that only by taking
the youngest reasonable value for the age of the globular clusters.
For currently favored values of $H_0$ we deduce that the SEC must
be violated somewhere between the formation of the oldest stars
and the present time.

Note the qualifications that should be attached to this claim: We
have to rely on both stellar structure calculations for $\tau_f$
and an estimate for $z_f$. Decreasing $z_f$ to be more in line with
the formation of the rest of the galactic structure ($z_f\approx
7$) makes the problem worse, not better ($H_0\leq54\pm7$ km s$^{-1}$
Mpc$^{-1}$). Increasing $z_f$ out to its maximum conceivable value,
$z_f\approx20$~\cite{Peebles3},  does not greatly improve
the fit to the SEC since the bound becomes $H_0\leq59\pm8$ km
s$^{-1}$ Mpc$^{-1}$.  All of these difficulties are occurring at
low cosmological temperatures ($T \leq 60$ K), and late times, in
a region where the basic equation of state of the cosmological
fluid is supposedly understood~\cite{eos}.

In contrast, the NEC does not provide any strong constraint on
$H_0$. For a spatially flat universe ($k=0$, $\Omega=1$, as preferred
by inflation advocates~\cite{Kolb-Turner,Linde,Moss})

\begin{equation}
\hbox{NEC}+(k=0) \implies \quad 
\tau = t_0-t \leq {\ln(1+z)\over H_0}.
\end{equation}

\noindent
Somewhat more complicated formulae can be derived for $k=\pm1$,
(open or closed universes).

This implies a (very weak) bound on $H_0$. In order
for cosmological expansion to be compatible with stellar evolution
and the NEC

\begin{equation}
\hbox{NEC}+(k=0) \implies \quad 
H_0 \leq {\ln(1+z_f)\over \tau_f}.
\end{equation}

The best value for $\tau_f$ ($16 \times10^9$ yr), and best guess
for $z_f$ ($z_f\approx15$), gives $H_0\leq 170$ km s$^{-1}$
Mpc$^{-1}$.  Decreasing $z_f$  to about $7$ reduces this bound
slightly to $H_0\leq 129$ km s$^{-1}$ Mpc$^{-1}$. Both of these
values are consistent with the observational bounds on $H_0$.  Even
for the highest Hubble parameter ($H_0=85$ km s$^{-1}$ Mpc$^{-1}$),
and oldest age for the oldest stars ($t_f = 18 \times 10^9$ yr),
$z_f\geq 3.6$ well within the observational bounds on $z_f$. The
present data is therefore not in conflict with the NEC.

The DEC provides us with a upper bound on the energy density $\rho$,
and therefore an upper bound on the rate of expansion. This translates
to a lower bound on the look-back time and a lower bound on the
Hubble parameter. For a spatially flat universe

\begin{equation}
\hbox{DEC}+(k=0) \implies \quad 
\tau = t_0-t \geq {1\over 3 H_0} {a_0^3-a^3\over a_0^3}.
\end{equation}

\noindent
Somewhat more complicated formulae can be derived for $k=\pm1$. In
terms of the red-shift

\begin{equation}
\hbox{DEC}+(k=0) \implies \quad 
\tau = t_0-t \geq {1\over 3 H_0} \left(1-{1\over(1+z)^3}\right).
\end{equation}

\noindent
So the ages of the oldest stars provide the constraint

\begin{equation}
\hbox{DEC}+(k=0) \implies \quad 
H_0 \geq  {1\over 3 \tau_f}\left(1-{1\over(1+z_f)^3}\right).
\end{equation}

\noindent
This also is a relatively weak constraint,

\begin{equation}
\hbox{DEC}+(k=0) \implies \quad 
H_0 \geq  20\pm3 \hbox{ km s$^{-1}$ Mpc$^{-1}$ }.
\end{equation}

\noindent
The present observational data is also not in conflict with the DEC. 


The estimated value of $H_0$ has historically exhibited considerable
flexibility: While it is clear that the relationship between the
distance and red-shift is essentially linear, the absolute calibration
of the slope of the Hubble diagram (velocity of recession versus
distance) has varied by more than an order of magnitude over the
course of this century. Hubble parameter estimates from 500 \hbox{
km s$^{-1}$ Mpc$^{-1}$ } to 25 \hbox{ km s$^{-1}$ Mpc$^{-1}$} can
be found in the published literature~\cite{Peebles5,Weinberg2}.
Current measurements give credence to the range 65---85 \hbox{ km
s$^{-1}$ Mpc$^{-1}$}~\cite{PDG}.  The reliability of the data on
$\tau_f$ and $z_f$ is harder to quantify, but there appears to be
broad consensus within the community on these
values~\cite{Peebles2,Peebles3,Peebles4}.

If the SEC is violated between the epoch of galaxy formation and
the present, then how does this affect our ideas concerning the
evolution of the universe? The two favorite ways of allowing SEC
violations in a classical field theory are by using a massive (or
self-interacting) scalar field~\cite{Hawking-Ellis3}, or by using
a positive cosmological constant~\cite{Visser4}.  A classical scalar
field $\varphi$, that interacts with itself via some scalar potential
$V(\varphi)$, can violate the SEC~\cite{Hawking-Ellis3}, but not
the NEC, WEC, and DEC~\cite{Visser4}. Indeed

\begin{equation}
(\rho+3p)|_\varphi = \dot\varphi^2 - V(\varphi).
\end{equation} 

\noindent
It is this potential violation of the SEC (depending on the details
of the time rate of change of the scalar field and its self
interaction potential) that makes cosmological scalar fields so
attractive to advocates of inflation~\cite{Kolb-Turner,Linde,Moss}.
In the present context, using a massive scalar field to deal with
the age-of-the-oldest-stars problem is tantamount to asserting that
a last dying gasp of inflation took place as the galaxies were
being formed. This is viewed as an unlikely scenario~\cite{last-gasp}.

In contrast, the current favorite fix for the age-of-the-oldest-stars
problem is to introduce a positive cosmological constant $\Lambda$~\cite{Leonard-Lake}, in which
case

\begin{equation}
(\rho+3p)_{total} = (\rho+3p)_{normal} - 2 \rho_\Lambda.
\end{equation}

\noindent
The observed SEC violations then imply

\begin{equation}
\rho_\Lambda \geq {1\over2}(\rho+3p)_{normal}.
\end{equation}

\noindent
Under the mild constraint that the pressure due to normal matter
in the present epoch be positive, this implies that more that 33\%
of the present-day energy density is due to a cosmological constant.


I have shown that high values of $H_0$ imply that the SEC must be
violated sometime between the epoch of galaxy formation and the
present. This implies that the age-of-the-oldest-stars problem
cannot simply be fixed by adjusting the equation of state of the
cosmological fluid. Since all normal matter satisfies the SEC,
fixing the age-of-the-oldest-stars problem will inescapably require
the introduction of ``abnormal'' matter---indeed large quantities
of abnormal matter, sufficient to overwhelm the gravitational
effects of the normal matter, are needed.


\null
\clearpage
\null





\begin{references}
\bibitem{look-back}
Look-back time to an object is simply defined as the difference
between the age of the universe now  and  the age of the universe
when the light that we are now receiving from that object was
emitted.
\bibitem{inflation} 
Cosmological inflation is a brief period of anomalously rapid
expansion in the early universe during which the universe inflates
by an enormous factor. Inflation is commonly invoked as a hypothesis
to explain the horizon problem, the flatness problem, and the
monopole problem as discussed in~\cite{Kolb-Turner,Linde,Moss}.
\bibitem{CC-absorbed}
A cosmological constant, if present, is absorbed into the definition
of the total density and total pressure.
\bibitem{Megaparsec}
Here I have expressed the Hubble parameter in terms of the standard
astrophysical units of kilometers per second per Megaparsec, with
a parsec being approximately $3\times 10^{16}$ m.
\bibitem{Peebles4}
See \cite{Peebles} p.~614. This number is a model-dependent estimate,
not an observation, fortunately the analysis of this report is
relatively insensitive to the precise value of $z_f$.
\bibitem{eos} The standard picture is that the universe is matter
dominated (i.e. dust) out to $z\approx 1000$, so that one expects
the equation of state to be $p=0$. See \cite{Peebles} p. 100.
\bibitem{last-gasp} 
Standard variants of inflation are driven by GUT-scale (grand
unified theory) phase transitions in the early universe and take
place when energies are of order $k T \approx 10^{14}$ GeV,
(see~\cite{Kolb-Turner}) with temperatures of order $T\approx 10^{27}$
K, whereas galaxy formation takes place for $T \leq 60$ K.
\end{references}
\end{document}